# Men-in-the-Middle Attack Simulation on Low Energy Wireless Devices using Software Define Radio

Mahyar TajDini[0000-0001-8875-3362], Volodymyr Sokolov[0000-0002-9349-7946], and Volodymyr Buriachok[0000-0002-4055-1494]

Borys Grinchenko Kyiv University, Kyiv, Ukraine
{m.tajdini,v.sokolov,v.buriachok}@kubg.edu.ua

**Abstract.** The article presents a method which organizes men-in-the-middle attack and penetration test on Bluetooth Low Energy devices and ZigBee packets by using software define radio with sniffing and spoofing packets, capture and analysis techniques on wireless waves with the focus on BLE. The paper contains the analysis of the latest scientific works in this area, provides a comparative analysis of SDRs with the rationale for the choice of hardware, gives the sequence order of actions for collecting wireless data packets and data collection from ZigBee and BLE devices, and analyzes ways which can improve captured wireless packet analysis techniques. The results of the experimental setup, collected for the study, were analyzed in real time and the collected wireless data packets were compared with the one, which have sent the origin. The result of the experiment shows the weaknesses of local wireless networks.

**Keywords:** Men-in-the-Middle, Software Define Radio, SDR, ZigBee, Bluetooth, BLE, Wireless Penetration Test, HackRF, GNU Radio.

## 1 Introduction

Capturing and analyzing Bluetooth traffic is very difficult and not cheap procedure comparing to alternative short-range wireless data networking technologies. However, the demand for affordable capture tools is increasing. The expansion of smartphones has provided to programmers the access to a huge built-in of Bluetooth-enabled computing platforms. Combined with recent updates to the Bluetooth standard, a new generation of Bluetooth-enabled devices now are capturing the market.

The development and testing of these peripheral devices and following smartphone applications are detained by inadequate or profusely expensive test tools. In order to improve the situation, this project report describes a realization an exhaustive Bluetooth Low Energy (BLE) capture tool based on GNU Radio software, and a usage of relatively cheap software-defined radio hardware.

With the spreading of cheap computers, high-speed data conversion devices and Software Define Radios (SDRs), a relatively cheap signal-processing platforms started to be available to individuals. The SDR application, which is going to be presented in this report, belongs to the cheap modern Bluetooth-enabled products.

This paper consists of the following parts: Section 2 "Related Works" contains the analysis of the latest scientific works in this area. Section 3 "Software Define Radio Fundamentals" provides a comparative analysis of the SDR and the rationale for the choice of hardware. Sections 4 "Bluetooth Low Energy Sniffing." Section 5 "ZigBee Experiment," and Section 6 "Multi-Channel Bluetooth Receiver" give the sequence order of actions for collecting wireless data packets and data collection from ZigBee and BLE devices. Ways to improve captured wireless packet analysis techniques are given in Section 7 "Capture and Analysis Techniques." The future work and plans are given in Section 8 "Conclusion and Future Work." The paper concludes with Sections "References."

## 2 Related Works

While we studied other scientific researches regarding SDR and low energy signal penetration we found out that the advantages of our point of view provides better and reliable results, comparing to other methods. In addition, we used the failure in IEEE 802.15.4 ZigBee networks to estimate the life cycle of the network in context of the network property [1] and made our stress check and man-in-the-middle attacks on the behalf of this.

As an alternative of the ZigBee radio modules there are nRF24L01+ radio modules that cost low but are powerful and extremely integrated. In addition, they use ultra-low power 2 Mbps RF transceiver for the 2.4 GHz philosophy ISM (Industrial, Scientific, and Medical) band. During this research, performances of nRF24L01+ modules were analyzed and compared upon the ZigBee modules in wireless ad-hoc networks [2]. We tried to capture information and find the highest performance by attacking our attack vector methods. In addition, we wanted to understand the hardware implementation in IoT networks [3], find an effective sniffing vulnerabilities [4] and to figure out the structure of work of low energy devices and their weak points.

## 3 Software Define Radio Fundamentals

SDR technology brings the pliability, price potency and power. In such way, they are pushing communication forward with wide benefits, accomplished by service suppliers and product developers through to end users. The Wireless Innovation Forum, which co-works with IEEE P1900.1 group, has tried to determine a definition of SDR that brings consistency and a clear summary of the technology and its associated benefits.

Traditional hardware based radio devices mostly limited by cross-functionality and can be changed just manually. This leads to higher prices of products and low flexibility in supporting multiple waveform standards. Opposite to this, SDR technology provides effective and relatively cheap solution for this question, permitting multi-mode, multi-band and/or multi-functional wireless devices that may be increased by usage of software upgrades. For each SDR (see Table 1) we have compared the price, frequency range, ADC resolution and maximum instantaneous bandwidth.

Table 1. SDR hardware comparison.

| Name | Cost, $ | Frequency range, MHz | ADC resolution, bits | Max bandwidth, MHz | Tx/Rx |
|---|---|---|---|---|---|
| R820T RTL2832U | 20 | 24–1766 | 8 | 2.4–3.2 | Rx only |
| Airspy R2 | 170 | 24–1750 | 12 | 10.0 | Rx only |
| HackRF One | 220 | 1–6000 | 8 | 20.0 | Tx/Rx (half duplex) |
| LimeSDR | 300 | 0.1–3800 | 12 | 61.4 | Tx/Rx |
| BladeRF | 650 | 300–3800 | 12 | 28.0 | Tx/Rx (half duplex) |

SDR platforms are Software and hardware toolkits that enable the designer to construct an SDR prototype or implementation. Of course, such implementation can include autonomous transmitter or receiver. Numerous SDR platforms are presented with different RF performances, programming languages and hardware architecture [5].

The frequency range grows up to 6 GHz with a DAC and ADC sampling rate up to 400 Msps (mega samples per second) and 100 Msps, correspondingly. Moreover, baseband process is handled though a GPP that executes a software toolkit, which the USRP develops through the named performances. Significantly, a GNU Radio has been developed to drive the USRP despite existing general software, like Simulink MATLAB and LabView. The GNU Radio is an open-source software, used to conduct not only the USRP, but another SDR hardwires too (e.g. HackRF and Nutaq ZeptoSDR [6]). Nevertheless, some communication standards and prototypes were enforced on GNU Radio and USRP (i. e. IEEE 802.15.4, IEEE 802.11a, IEEE 802.11p, automatic dependent surveillance-broadcast and high definition TV, etc.). As for the SDR based USRP and GNU Radio, they will be described further more deeply.

Due to all radio devices are different in effectiveness at handling frequency ranges and signal types, penetration testers should be sure that the device meets their RF spectrum features. As a result, we found out that the HackRF One is the best choice for this experiment, because: it has strong features, but the most important is a good support by different open-source software on most standard computer platforms; it costs almost $220 and brings more capabilities than lower cost devices; it supports wide Frequency range and up to 20 Msps.

## 4  Bluetooth Low Energy Sniffing

We created an iBEACON packet from phone, and a web URL as a packet data on channel 37. After that, we have broadcast it and then run it out ". /btle_rx –g 0":

```
163342us Pkt192 Ch37 AA:8e89bed6
ADV_PDU_t2:ADV_NONCONN_IND T1 R0 PloadL25
```

```
AdvA:41e0302e6669
Data:0303aafe0e16aafe10bb0074616a64696e690a CRC0
```

Definitely, the iBeacons are a trending topic nowadays. They allow indoor placing, while letting your phone to know that you are in range of a beacon. This can allow different usage: from helping you to search out your car in a parking garage, through coupons and location-aware special offers in retail, to a full ton of apps that we can't imagine right now.

The main benefit in BLE is, of course, the low energy consumption. For example, some beacons can spread a signal during two years on a single cell battery (the batteries mostly are not replaceable. It is possible just to change the beacon when it stop working). We should admit that both "classic" and BLE use the same spectrum range 2.4000–2.4835 GHz. The BLE protocol has lower transfer rates, but it is not important for data stream, while for discovery and simple communication it is.

In terms of range, both BLE and "classic" Bluetooth signal can reach up to 100 m.

## 5   ZigBee Experiment

In our experiment, we used Pololu Wixel as ZigBee Rx/Tx module to send one packet per second with static data as we programmed it in our case FFEEFFEE on 2.4499 GHz.

The Pololu Wixel is a general-purpose programmable module that includes a 2.4 GHz radio and USB. The Wixel relies on the CC2511F32 microcontroller from Texas Instruments, that works on 2400.0–2483.5 MHz frequency range.

As mentioned within the ZigBee protocol stack, the ZigBee MAC layer frame is composed of MAC header, MAC payload and FCS. This part is also points toward MAC Protocol Data Unit (MPDU) and sets into Physical Packet Data Unit (PPDU) frame of ZigBee. We've chosen Man-in-The-Middle Attack to capture data, to resend it again as Spoofed source and to show how a middle-man here as a device can spoof and change data in between by reassembling data and resend it to destination again while destination device cannot detect it. HackRF has a role as our middle-man device. Without cutting the signal, we can send and receive packets with approximately 7% of errors as it shown in Fig. 1. Then we tried to catch these signals with HackRF and resend them by cloning the packets. For this purpose the GNU Radio was used, which have to catch and save the data in binary file.

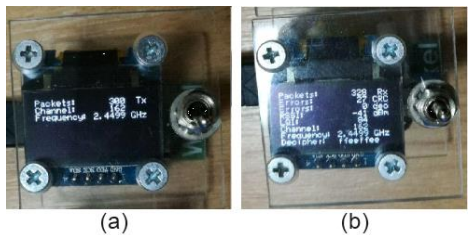

**Fig. 1.** Experimental packet transmitter (a) and receiver (b).

The result shows: once sink file are going to be "File" path with the name of "rec1" as we can see in Fig. 2. Then we have created a transceiver to send the same signals as we had caught in binary file. The scheme of this transceiver is shown in Fig. 3.

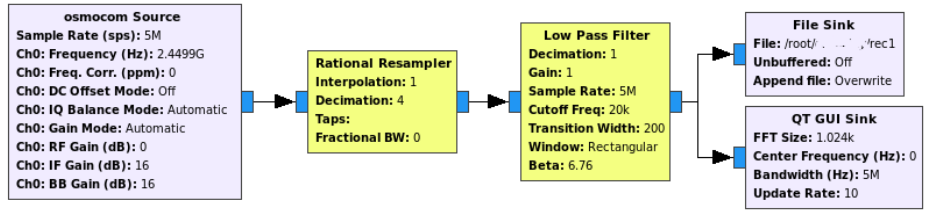

**Fig. 2.** Receiver block in GNU Radio.

In this experiment, we have received more than 50% of signals without error. It means that in real communication, we can have just 7% of packet lost and after its sniffing by HackRF and while resending we have 43% of packet lost. This is a very good statistic for such experiment. The obtained information gives possibility to jam signals for a few times and catch, for example, authentication packets. Then it can be resent to authenticator. In another experiment, we caught a car remote control signals, which had been on 433 MHz range. By same ratio, we could open and close the car door without real remote control key.

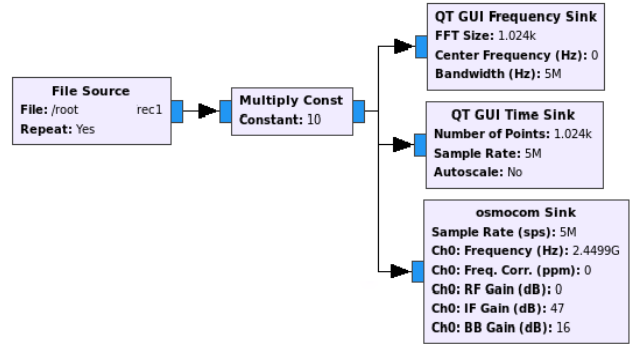

**Fig. 3.** Transceiver block in GNU Radio.

## 6 Multi-Channel Bluetooth Receiver

The first part of the current project represents a capture of a portion of the ISM band to a file, with post-processing usage of the GNU Radio framework. The implementation is revealed on the Internet [7] under open-source license.

### 6.1 Sample Acquisition

In Fig. 4 is showed the signal flow applied by the USRP hardware throughout a capture session. The RFX2400 and HackRF module covers the analog portion from the antenna, and therefore the USRP N210 main-board clothes the process to the sample capture file. The RFX2400 and HackRF module includes a SAW filter, which limits the receiver spectrum to the 2.4 GHz band. RF mixed with a frequency-agile quadrature local oscillator, choose the center of the pass-band. The quadrature mixer includes an anti-aliasing low-pass elliptical filter. The USRP samples are the quadrature analog signals to 14-bit digital samples at 100 Msps. Carried out in the USRP FPGA digital down-converter provides a decimated stream at a selectable rate. In this research a 25 Msps capture is provided as an example. The capture stream is transferred across the USRP's gigabit Ethernet port to a PC, which records the sample series to a capture file.

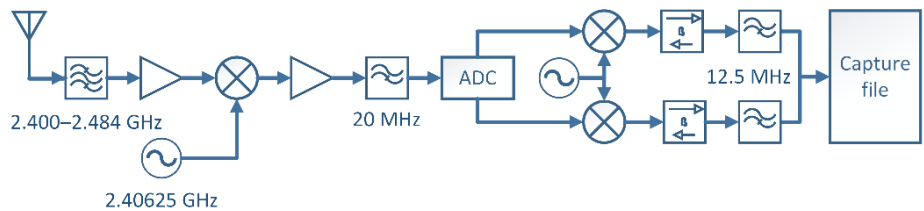

**Fig. 4.** Sample acquisition diagram.

### 6.2 Packet Recovery

This section provides a review of the BLE packet recovery, performed in the first stage of this research. More detailed descriptions of the assorted signal processing of sub-blocks you can see further. The implementation is accomplished by applying to the existing modules from the GNU Radio toolkit and new modules, developed exactly for this research. BLE packets, recovered with a software post-processing stage, applying to the capture file. The signal process flow is shown in two major elements: the symbol recovery for the BLE channels represented by the capture file and the symbol decoding and squelch operations (Fig. 5).

BLE packet recovery goes over all RF channels covered by the capture file. A frequency-translating, decimating FIR filter provides a 2 Msps sample stream for the channel with a 3 dB bandwidth of 500 kHz, which is shown at point *C*. At the same time, a FIR band-pass filter provides a sample stream, which offsets on 790 kHz from the channel center, shown at point *B*. After that, the on-channel samples are converted to symbols (phase differences) with a differential demodulator. A Mueller & Müller Clock-Recovery block aligns the demodulated symbols to the center that is why the final hard-decision block provides recovered symbols at point *A*. The data recovered at points *A*, *B*, and *C* can be detected when a valid Bluetooth low-energy packet is available. Also the packet-presence qualification includes both: a signal-energy squelch component and sequence-matching component.

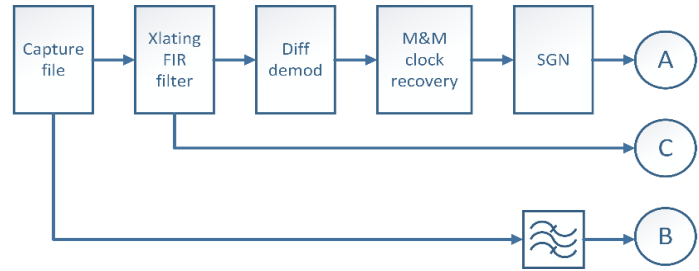

**Fig. 5.** Bluetooth image recovery and off-channel band pass filter.

At the same time, the recovered symbol stream is compared with the fixed portions of the BLE packet format. All BLE packets should start with an 8-symbol preamble (PA) that consists of alternating 1's and 0's. The 32-symbol access address (AA) is fixed for symbol streams, which are recovered from advertising channels. Therefore it is used in the decode decision. The 16-symbol de-whitened by PDU header is inspected also and compared with possible candidates. The decode stage leads to a matching metric, primarily a Hamming distance, computed for the candidate packet. If this metric exceeds a programmable threshold, then the symbol stream can be interpreted as a complete packet, and admitted to the recovery log (Fig. 6).

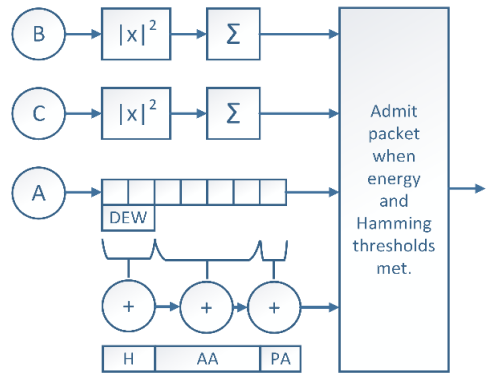

**Fig. 6.** Bluetooth low energy squelch and image decode.

### 6.3   Frequency-Translating Decimating FIR Filter

The frequency-translating decimating FIR filter is applied to the sample stream for any channel with a purpose to be covered by the sampling spectrum. All process proceeds with floating-point complex values, where both input and output streams apply in quadrature. The project implementation uses an existing element from the GNU Radio toolkit. The FIR filter is low-pass with a 3 dB cutoff of 500 kHz and transition width of 300 kHz, modelled with a Hann Window 8, yielding 355 taps. The decimation rate is chosen to bring a 2 Msps output, based on the input sampling rate. Consequently, 2 Msps is the lowest input sampling rate supported by this research.

### 6.4 Band-Pass Noise Filter

The second frequency-translating, decimating FIR filter is applied to the input sample stream to feed the SNR-based squelch mechanism. The FIR filter is the low-pass with a 3 dB cutoff of 22.5 Hz and a transition bandwidth of 10 kHz, modelled with a Hann window, yielding 7751 taps. The filter calls to a pseudo-carrier that is 790 kHz offset from the on-channel carrier. After that the filter provides an equivalent decimation rate like the on-channel filter and leads to a 2 Msps complex sample stream.

### 6.5 Differential Detector

Symbols are picked out as phase shifts, which are determined by a differential demodulator operating. The project implementation modifies an existing component from the GNU Radio toolkit. Complex input samples are processed in quadrature. The resulting symbols of the recovered soft should be a floating-point values. Because the decimation is not performed, the output symbol rate is 2 Msps.

### 6.6 Mueller & Müller Clock Recovery

The clock recovery determines and tracks a fractional offset between samples, which represent the peak excursion of the (interpolated) recovered symbols. Such information calls to remained, that previously we mentioned about the project implementation, which modifies an existing component from the GNU Radio toolkit. The 8-tap FIR filter performs interpolation and permits the shape of the recovered symbol pulses to be approximated. Also the interpolated value is represented as the soft symbol at the clock recovery point. In addition, the output of this operation decimates by a factor of two, which results a 1 Msps symbol rate. Therefore, the floating-point soft symbols simply apply to a hard-decision mapper to yield a recovered hard symbol stream at 1 Msps. This corresponds to the point *A* in Fig. 5 sample acquisition diagram [8].

### 6.7 Squelch Operation

In the previous paragraph, we describe the squelch appeal. The implementation was developed exactly for this research. The energy in the on-channel samples and off-channel samples is determined by summation the magnitude-squared for all samples in the decode window. If the on-channel energy exceeds a given threshold, and the ratio between on-channel and off-channel energies reaches its threshold, then the squelch asserts and permits the decoder to work.

### 6.8 Symbol Decoding

The symbol decoding works as described in the previous paragraph and the implementation was developed specially for this research. The decoding continues qualified by the squelch operation, what was mentioned above. A 56-symbol candidate decode window is processed every time when the decoder operates. The first 40 symbols work up as received, and after that the next 16 symbols, which represent the PDU header, will be de-whitened before further process. Should be admitted that the 56-symbol window is usually compared with the fixed portions of the BLE packet, specifically with the preamble and fixed-zero portions of the PDU header. A Hamming distance calculation is performed crucial base matching metric for all candidate packets.

Then the matching metric increases by a representative value of the access address match. The Hamming adds distance to the fixed access address for advertising channels. As for data channels, the matching metric increases by the amount of offenses to the AA limitations. The decoded packets can be admitted to the output when the matching metric is less than a threshold. In this project implementation any matching metric, which is less than 3, permits the decoder to proceed and recover the whole packet. The remaining packet symbols are de-whitened too before the admission.

### 6.9 Example Results

To verify the correct operation of the project implementation we used few real-world captures. Three commercial BLE devices were accustomed to generate the radio transmissions, which were necessary for testing. The used devices were Apple iPad mini, which operates as BLE scanner and initiator$; Polar H7 heart monitor operates as BLE advertiser; Texas Instruments CCTAG operates as BLE advertiser. Two sample runs are represented here. The captures related to these sample runs were used to validate the signal process blocks, which are described above.

Each run consists of a one-second window of RF samples, collected at 25 Msps and focused on 2406.25 MHz. The USRP introduces with the software that adds the generic RF streaming to a capture file. During every capture, iPad was obtaining the instructions to establish a connection with one of the mentioned devices. The results of the each capture are listed in Table 2.

**Table 2.** Bluetooth low energy packet capture and recovery examples.

| Device | False-positive packet count | Recovered packet count | Reported SNR, dB |
|---|---|---|---|
| Polar H7 | 37 | 27 | 22.3 |
| TI CCTAG | 33 | 5 | 34.7 |

## 7 Capture and Analysis Techniques

The represented project applies wide-band capture techniques and performs analysis of the captured BLE packets. The most valuable BLE packet exchanges appear on data channels when devices are in the Connection state, due to needs of the research, product development and security analysis. Because the capture and analysis session can start with BLE devices in randomly, we should be ready to situation when the captured traffic covers already connected devices.

Whenever a CONNECT_REQ is analyzed, the connection data can be used for ongoing traffic analysis. While the data packets are available, certain techniques can be used to detect the parameters. We should admit that the wide-band approach with its exhaustive capture does not require any connection parameters out of the access address as long as RF conditions are good. However, if the RF conditions are poor, the missed packets cannot be predicted without known Interval, Hop and ChM parameters. The current project is intended to incorporate the capture of valid packets, invalid packets to a particular tolerance and a recognition of missing packets.

## 8 Conclusion and Future Work

In parallel and continuously we are working on SDR implementations for IEEE 802.15.4-based wireless detector network that can be a set of detector nodes, which communicate throughout frequency links. This simplified definition primarily involves three lowest layers of OSI model. Such layers are physical data link and network layers. Current analysis, in most cases, touches separately every issue of these layers. We focus on software transmitter/receiver for 868/915 MHz PHY and cognitive wireless sensor network based on IEEE 802.15.4, which scientifically with mathematical formula and experiment statistic shows vulnerabilities and attacks with the help of SDR.